\documentclass{aa}  
\usepackage{graphicx}
\usepackage{natbib}
\usepackage{txfonts}
\usepackage{color}
\usepackage{rotating}
\usepackage{longtable,lscape}
\begin{document}
   \title{New white dwarfs in the Hyades}
\subtitle{Results from kinematic and photometric studies} 
\author{E.~Schilbach  \and
S.~R\"{o}ser   }

   \offprints{S.~R\"{o}ser}

\institute{Astronomisches Rechen-Institut, Zentrum f\"ur Astronomie der Universit\"at Heidelberg, M\"{o}nchhofstra\ss{}e 12-14,
D--69120 Heidelberg, Germany\\
email: elena@ari.uni-heidelberg.de, roeser@ari.uni-heidelberg.de
}
   \date{Received July 12, 2011; accepted November 14, 2011}

 
  \abstract
  {} 
   {On the basis of the PPMXL catalogue (R\"oser, Demleitner \& Schilbach, 2010) we searched for white dwarfs
   that are also member
   candidates of the Hyades in a region up to 40~pc from the cluster centre.}
   {We used the proper motions from PPMXL in the convergent
point method to determine probable kinematic members. We cross-matched the kinematic candidates
with catalogues containing white dwarfs 
and, finally, checked the kinematic with the photometric distances for consistency.}
   {We found the 10 classical white dwarfs in the Hyades and determined their individual
kinematic distances. Additionally, we identified 17 new probable (former) Hyades white dwarfs,
i.e. white dwarfs co-moving with the bulk space motion of the Hyades cluster. At present, none of them
can be excluded from membership on the basis of the
measured radial velocities.
For another 10 objects, the kinematic and the photometric distances disagree,
which rates them as probable non-members. Among the probable members, five white dwarfs are in binary systems,
three are known, two are new. There is good indication for an empirical magnitude-distance (from centre) relation,
such that the dimmer white dwarfs are farther away from the cluster centre than the brighter ones. 
Our sample becomes incomplete close
behind the centre of the cluster. 
Follow-up observations are encouraged to independently confirm the
predicted radial velocities and the distances of the candidates.}
   {}
  \maketitle
%
\section{Introduction}
The 650 Myr old Hyades cluster is, in many respects, a test case for theories
of stellar evolution in, and dynamical evolution of, open clusters in our Galaxy.
The white dwarfs, in particular, shed light on the high-mass end of the initial mass function (IMF)
in open clusters and the subsequent development of the cluster's mass function.
In an influential paper, \citet{1992AJ....104.1876W} have analysed the white dwarf
population in the Hyades, and concluded that the cluster should contain 
at least 21 white dwarfs
dimmer than the 7 confirmed white dwarf members known at that time. The authors derived this
finding from adopting a Salpeter IMF normalised via the 24 brightest
main-sequence stars presently residing in the Hyades. From different
considerations, \citet{1988AJ.....96..198G} arrived at a number between 50 and 150 white dwarfs
originally present in the cluster. On the other hand, \citet{1998AJ....115.1536V}
counted only 10 white dwarfs as known members in the Hyades when he investigated the contribution
of white dwarfs to the masses of open clusters.

To summarise, not much progress has been made to solve the discrepancy between the number of white dwarfs
estimated from the IMF and the actual number of white dwarfs found in the Hyades.
\citet{1992AJ....104.1876W} mention in their paper that white dwarfs initially present in the bound 
Hyades have left the cluster in the meantime, and the authors suspect them in the Hyades supercluster as defined by
\citet{1958MNRAS.118...65E}. \citet{1992AJ....104.1876W} checked the \citet{1999ApJS..121....1M} catalogue by analysing the space motions of the white dwarfs therein, and claim to
have detected a handful of stars moving within 13 degrees from the Hyades convergent point and tangential velocities within
$\pm 2$ $\rm{km\,s}^{-1}$ of the cluster motion. Extending the search to $\pm 5$ $\rm{km\,s}^{-1}$, they found that 
about 2/5 of all nearby white dwarfs may be related to the Hyades.
Unfortunately,  \citet{1992AJ....104.1876W} did not publish the data of their candidates,
therefore we could not compare them with our results below. As a mechanism
for white dwarf loss,
\citet{2003ApJ...595L..53F} proposed that white dwarfs could be expelled from their parent cluster
through non-spherically symmetric mass loss during the post-main-sequence evolution, which leads to a
recoil speed of a few kilometres per second for the white dwarf remnant.

\citet{1992AJ....104.1876W} also discussed the possibility that missing 
white dwarfs can hide themselves behind the red dwarf
companion in binary systems. They estimated that the missing ones could only be found, even in the $B,V$ bands,
if the other component is later than spectral type G.

In a previous paper \citep[][henceforth Paper I]{2011A&A...531A..92R} we have analysed
the Hyades cluster and their surroundings up to 30 pc to search for
main-sequence stars as member candidates. We found that the present-day tidal radius
is about 9 pc, and 275~$M_\odot$ (364 stellar systems) are gravitationally bound.
Outside the tidal radius we found
another 100~$M_\odot $ in a volume between one and two tidal radii (halo),
and another 60~$M_\odot $ up to a distance of 30 pc from the centre. From their
kinematics we infer that the stars outside the tidal radius are formerly bound members that
left the cluster. It is therefore appropriate to repeat a selection process similar to that
in Paper I to search for white dwarfs up to 30 pc and more from the cluster centre. Compared
to earlier studies of Hyades white dwarfs,
which performed deep searches in limited fields-of-view, we have the advantage to be able to use the
deep all-sky astrometric survey PPMXL \citep{2010AJ....139.2440R}.

This paper presents candidates that have five out of six phase space parameters compatible
with their Hyades origin. Once available, their true space motion in radial direction must confirm or reject them.
Our list of candidates from Table \ref{table:1} can serve as an input catalogue for future observations.
Therefore we are hesitant at this stage to draw far-reaching conclusions, e.g. on the IMF of the Hyades,
on the problematics of 
cooling ages, or on the initial mass-final mass relation.

The paper is structured as follows: after a short listing of the so-called ``classical'' Hyades
white dwarfs in Sect. \ref{classical}, we describe our
selection process in Sect. \ref{select}. Section \ref{individual} follows with comments to each individual
candidate. The questions of spatial distribution and completeness
of the sample are covered in Sect. \ref{complete}. Finally, the discussion in Sect. \ref{discus} completes the paper.
\section{The ``classical'' Hyades white dwarfs}~\label{classical}
The paper by \citet{1998AJ....115.1536V} lists 10 white dwarfs that we call henceforth the ``classical'' Hyades
white dwarfs. There are seven single white dwarfs and three in binary systems. We list them with their
primary identifiers in SIMBAD (Data base of the Centre de Donn\'ees astronomiques de Strasbourg, CDS).
The seven single white dwarfs are EGGR 26, 29, 36, 37, 39, 42, and 316.
The three stars in binary systems are HR 1358, EGGR 38, and V471 Tau.
Data for these stars are given in Table~\ref{table:1} (the first ten rows).
In some cases  there is no consistency about actual membership in the cluster.
For instance, \citet{1992AJ....104.1876W} exclude EGGR 29 from membership, because they put it at 60 pc, where its tangential
velocity would be discordant from the bulk tangential velocity, whereas \citet{2009ApJ...696...12D} used it for their
determination of the white dwarf age of the Hyades.


\citet{1992MNRAS.257..257R} lists two additional candidates RHya 102 and RHya 145. He also
examined all candidates for Hyades white dwarfs proposed by  \citet{1969AJ.....74....2V}, and
discarded all of them except vA54 and vA71, which were outside his field-of-view. We discuss
these four objects  in Sect. \ref{individual}.   

Throughout the paper we use the following abbreviations for star names:
   \[
      \begin{array}{lp{0.8\linewidth}}
\rm{VR} & \citet{1934MNRAS..94..508V}\\
\rm{HZ} & \citet{1947ApJ...105...85H}\\
\rm{HG7} & \citet{1962LowOB...5..257G}\\
\rm{HR} & Catalogue of bright stars, \citet{1964cbs..book.....H}\\
\rm{EGGR}& \citet{1965ApJ...141...83E}\\
\rm{vA} & \citet{1969AJ.....74....2V}\\
\rm{RHya} & \citet{1992MNRAS.257..257R}\\
\rm{WD} & \citet{1999ApJS..121....1M}, updated version 2008 \\
\rm{LB} & (Luyten, Blue), Luyten W.J., Various lists published by Luyten under the general title:
 A Search for Faint Blue Stars (50 papers)\\
\rm{LP} & (Luyten, Palomar obs.), Published from 1963 to 1981 in Univ. Minnesota, Minneapolis, fascicules 1 to 57\\
\rm{GJ} & CNS3, Catalogue of Nearby Stars, \citep{1991adc..rept.....G}.
      \end{array}
   \]
 \begin{figure*}[t!]
   \centering
   \includegraphics[bb=76 42 465 785,angle=270,width=17cm,clip]{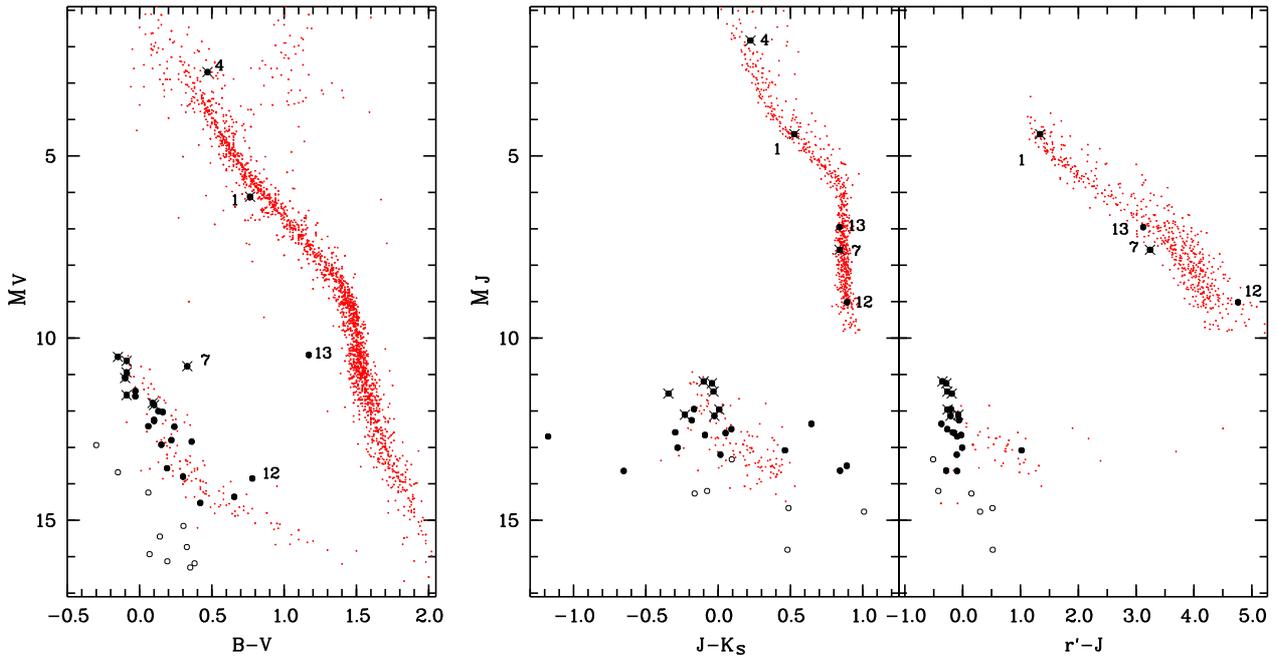}
      \caption{Colour-absolute-magnitude diagrams: $ M_V $ vs. $ B-V$ (left), $M_J $  vs. $J-K_s$ (middle)
and $ M_J  $ vs. $  r'-J$ (right). In all diagrams the black dots show the probable Hyades white
dwarfs from this paper, the 10 classical white dwarfs are marked additionally with crosses.
The black circles show also spectroscopically confirmed white dwarfs from \citet{1999ApJS..121....1M},
which are probably non-Hyades though. The small (red) dots in the left panel show the main and
the degenerate sequences of stars from the CNS3 \citep{1991adc..rept.....G}. In the middle and the right panels
the degenerate sequences are again from CNS3, while the 724 Hyades members from Paper I
represent the main sequences. Note that the brightest stars have no $r'$ magnitudes (right panel)
in CMC14. Spectroscopic binaries are marked by their numbers in Table~\ref{table:1}.}
         \label{cmd}
   \end{figure*}
\section{The selection process}~\label{select}
In Paper I we described in detail how we selected MS (main sequence) Hyades candidates from their
kinematic and photometric properties in
the PPMXL catalogue \citep{2010AJ....139.2440R}. More specifically, we used the Carlsberg-UCAC (CU) subset containing
improved proper motions and photometry by including
UCAC3 \citep{2010AJ....139.2184Z} and CMC14, Carlsberg Meridian Catalog 14, \citep{2006yCat.1304....0C}.
Paper I also contains a description of the convergent point method, which served as a baseline for the selection.
For our general selection process we did not permit the tangential motion to differ
from the one given by the bulk
motion of the cluster by more than 4 $\rm{km\,s}^{-1}$,
and we considered only stars closer than 30 pc from the cluster
centre. 
The stars had to be in the CU subset to PPMXL to ensure that they have CMC14 and/or 2MASS  
\citep{2006AJ....131.1163S} measurements.
These kinematic selection criteria were fulfilled by 15757 stars out of the 140 million contained
in the CU subset, which were
shown in Fig.~1 of Paper I. 
Two features from the kinematic selection via the convergent point method are worth to be explained in
more detail. First, for each candidate that is supposed to share the bulk space motion of the cluster,
the convergent point method predicts
a radial velocity that only depends
on the position $\alpha, \delta$ of the star. In consequence, the radial velocity of each candidate has to be
measured to confirm it is still member.  
Second,
the convergent point method attributes (predicts) a so-called secular parallax to each candidate by
minimising the difference between the space motion of the cluster centre and the space motion of a candidate star.
In a very strict sense the secular parallax is not the least-squares solution in the plane perpendicular to the
line-of-sight, but this difference is small in all practical cases. So, the second confirmation of
membership comes from an independent measurement of the distance to the star.
If the result of this independent
distance determination does not confirm the prediction, this means in turn that the difference in
space motion between the bulk of the cluster and the candidate increases. This may rule out a given star
as a member candidate.

As an independent check of the predicted distances of kinematic candidates, we can consider their location in
different colour-absolute-magnitude diagrams. In Paper I we used $B,V$-photometry for bright stars;
for stars fainter than $M_{K_s} = 4$ the basic photometric data were
$r'$ from CMC14 and $JHK_s$. Only for 724 out of 15757 kinematic candidates, the membership was
confirmed by photometric selection in Paper I.

The procedure adopted in Paper I is a two-step procedure. The first step is the kinematic
selection, and the second is the check of kinematically predicted distances with photometric ones, i.e.
a comparison with isochrones.
In principle, the same approach could be applied to find 
white dwarfs among kinematic candidates. This would require an all-sky, accurate, multi (at least two)-colour
photometric survey. Concentrating on a 30 pc radius around the centre of the Hyades, such a survey should
at least cover 12.1\% of the celestial sphere or 5000 square degrees. The only survey that fulfils this requirement
is 2MASS in the near-infrared.
However, even 
at the central distance of the Hyades (46.3 pc), white dwarfs will be at the limiting magnitude
of 2MASS and also of CMC14, where the photometric quality becomes very poor.
Therefore, we introduced an intermediate step, in which all kinematic candidates from Paper I were cross-matched with the
white dwarf catalogues by Luyten (VizieR Online Data Catalog III/70) and \citet{1999ApJS..121....1M},
the updated version (VizieR Online Data Catalog III/235B). Each match was individually 
checked using the VizieR data base from the CDS. After rejecting obvious red dwarfs (from Luyten candidates),
we identified 20 white dwarfs that passed the kinematic selection, among which were
all 10 classical Hyades white dwarfs.
Ten more stars would be added if the kinematic criteria were relaxed to 40~pc and 5 $\rm{km\,s}^{-1}$.
Finally, we found six faint white dwarf candidates in the  PPMXL (i.e. without CMC14 and/or 2MASS measurements).
One more white dwarf was found in the Prosser \& Stauffer data base
(presently available from J. Stauffer, priv. comm.), although it is not explicitly mentioned there as a
white dwarf. For this star, No. 20  in Table \ref{table:1},
PPMXL gives wrong proper motions, therefore we took them from \citet{2006A&A...448.1235D}.
In total, the sample of the Hyades white dwarf candidates contains 37 stars. All but three of them
(12, 15, 30) are spectroscopically confirmed white dwarfs.  

For these 37 stars we then examined if their loci in colour-absolute magnitude diagrams are consistent with
those of white dwarfs. As mentioned above, 2MASS and CMC14 are not well suited for the fainter candidates.
Unfortunately, the UKIRT Infrared Deep Sky Survey \citep[UKIDSS,][]{2007MNRAS.379.1599L} provides no help for this problem, because 
in the region of the Hyades it is constrained to the central 292 square degrees, and only K-band photometry
is  available for these. Indeed, except for the 10 already known white dwarfs, only stars No. 20, 21, and 22
even have K-band photometry in UKIDSS. 
Since the vast majority of stars in the PPMXL and its CU subset do not have 
appropriate photometric data in the optical, 
we took the $B, V$ magnitudes for the Hyades white dwarf candidates  
from the VizieR data base (CDS). The sources are given as footnotes to Table~\ref{table:1}. 
For three white dwarfs (entries 28, 29, 35 in Table~\ref{table:1}),
no original $V$ measurements are available. In \citet{1999ApJS..121....1M} they are SDSS white
dwarfs, and we converted their $ugriz$ magnitudes into $V$ and $B-V$ using the transformations
from \citet{2005AJ....130..873J}. Also, all 37 stars have been visually inspected on the
digitised sky survey charts from IRSA (NASA/IPAC Infrared Science Archive) to avoid
coarse misidentifications of the different surveys.
 
In the following paragraphs we check if the kinematically selected stars populate allowed loci
in the colour-absolute-magnitude diagrams (CMDs). In Fig. \ref{cmd} we show three CMDs,
($M_V $  vs. $  B-V $, left), ($M_J $  vs. $  J-K_s$, middle) and
($M_J $  vs. $ r'-J$, right). As references for the loci of the degenerate stars we have taken
the white dwarfs from the CNS3 in all three panels (small red dots). For main sequence dwarfs the loci
of the 724 Hyades from Paper I (also small red dots) are taken in the middle and right diagrams.
Because in most cases we could not find precise B and V magnitudes for our sample of stars from Paper I,
we took the main sequence in the left panel again from the CNS3.

All 37 stars discussed in this paper are marked in black in Fig. \ref{cmd}. 
The field white dwarfs from the CNS3 show a relatively well-defined sequence in the optical
(the left diagram), and a number of our Hyades white dwarf candidates follow this sequence.
We conclude that the convergent point method has correctly predicted the distances for these stars and
refer to them as probable Hyades members marked as solid dots in Fig. \ref{cmd}. 
The classical 10 Hyades white dwarfs are additionally indicated by crosses.
In this panel we find 10 white dwarfs, marked by open circles
which would be sub-luminous if set at their kinematic distances. We conclude that these stars must be
at farther distances from the Sun than predicted, therefore have higher tangential velocities,
and must be rated as probable non-Hyades.
The best coincidence between the probable Hyades white dwarfs
and the reference loci is seen in the left panel, one of the reasons being the better optical photometry
available for the white dwarfs in the Johnson $B,V$ system. In the  near-infrared, NIR, panel (middle) the scatter in the
$  J-K_s$ colour is much larger than in $  B-V $, because the fainter stars are at the
detection limit of 2MASS, especially in the $K_s$ band. To a moderate extent, this holds 
for the  $ r'-J$ colour, too.

We find five stars in the middle panel ($M_J $  vs. $  J-K_s$)
that perfectly lie on the Hyades NIR main sequence. 
We marked these stars in Fig. \ref{cmd} with their numbers from Table~\ref{table:1}. 
All five are included in our sample of 724 Hyades members in Paper~I. 
Four of them (1, 4, 7 and 13 or HR 1358, EGGR 38, V471 Tau and 
WD 0217+375, respectively) are known as binaries 
containing a WD and a MS component. The first three belong to the classical Hyades members,
whereas WD 0217+375 (13) has not be associated with the Hyades before.
LP 649-0071 (12) is rated a white dwarf in Luyten's White Dwarf Catalogues,
though it has a NIR-colour typical of red dwarfs.
This indicates a possible binary nature of this object. We discuss its properties in more detail in section
\ref{individual}.
\begin{table*}
\centering
\caption{Hyades white dwarf candidates.} 
\label{table:1}
\begin{tabular}{r c lrrrrrrrrc c c}        
\hline\hline                 
 Star  & other name & SpT & D & r$_c$ & v$_\perp$ & $M_V$ & $B-V$ & $M_J$ & $J-K_s$ & $r'-J$& X &Bin.&Ref. \\ 
 No. & & &   [pc] & [pc] & $[\rm{km\,s}^{-1}]$ &[mag] &[mag]  &[mag]  &[mag] &[mag] & & &$B,V$\\   
\hline
  1 &   V471 Tau          &DA1.5 & 47.3 &  7.5 &  0.1 &  6.12 & 0.76 &  4.40 & 0.53 & 1.34 &y &y &1,1 \\
  2 &   EGGR 26, HZ 4     &DA4   & 35.1 & 13.5 &  0.9 & 11.83 & 0.10 & 12.11 &-0.23 &-0.08 &y &- &2,2 \\
  3 &   EGGR 29, LB 227   &DA3   & 51.7 &  6.6 & -0.4 & 11.78 & 0.09 & 12.13 &-0.03 &-0.21 &y &- &3,3 \\
  4 &   HR 1358           &DA3   & 49.1 &  3.9 & -0.2 &  2.70 & 0.47 &  1.83 & 0.22 &  --- &y &y &1,1 \\
  5 &   EGGR 36, VR 7     &DA2.5 & 44.1 &  2.4 & -0.8 & 11.10 &-0.10 & 11.53 &-0.34 &-0.19 &y &- &2,2 \\
  6 &   EGGR 37, VR 16    &DA    & 47.7 &  1.5 & -0.5 & 10.62 &-0.09 & 11.23 &-0.04 &-0.28 &y &- &2,2 \\
  7 &   EGGR 38, HZ 9     &DA2.5 & 43.2 &  3.3 & -0.9 & 10.78 & 0.33 &  7.58 & 0.84 & 3.24 &y &y &2,2 \\
  8 &   EGGR 39, HZ 7     &DA2.3 & 45.7 &  3.4 &  0.3 & 10.94 &-0.09 & 11.47 &-0.03 &-0.27 &- &- &2,2 \\
  9 &EGGR 316, LP 475-242 &DB4   & 46.9 &  3.3 & -0.7 & 11.56 &-0.09 & 11.96 & 0.01 &-0.26 &- &- &2,2 \\
 10 &   EGGR 42, HZ14     &DA2   & 46.0 &  5.2 &  0.2 & 10.51 &-0.15 & 11.18 &-0.10 &-0.35 &- &- &2,2 \\
\hline 
 11 &   WD 0120-024       &DC    & 39.0 & 36.3 &  0.8 & 14.52 & 0.42 & 13.50 & 0.89 &  --- &- &- &7,4 \\
 12 &   LP 649-0071       &---   & 34.3 & 29.1 & -1.3 & 13.85 & 0.78 &  9.01 & 0.89 & 4.76 &y &? &8,5 \\
 13 &   WD 0217+375       &DA    & 25.1 & 29.6 &  0.6 & 10.46 & 1.17 &  6.95 & 0.84 & 3.12 &- &y &7,7 \\
 14 &   WD 0230+343       &DA    & 37.8 & 24.1 &  0.8 & 12.84 & 0.36 & 13.63 & 0.84 &-0.29 &- &- &6,6 \\
 15 &   LP 246-0014       &---   & 35.1 & 23.5 &  2.4 & 13.80 & 0.30 & 13.65 &-0.65 &-0.10 &? &- &8,7 \\
 16 &   WD 0259+378       &DA3   & 66.0 & 33.6 & -4.0 & 11.44 &-0.03 & 12.35 & 0.64 &-0.37 &- &- &3,3 \\
 17 &   WD 0312+220       &DA2.5 & 44.4 & 14.3 &  4.7 & 12.43 & 0.24 & 13.01 &-0.28 &-0.01 &y &- &7,7 \\
 18 &   LP 653-0026       &DA3.5 & 28.5 & 23.0 & -1.9 & 12.93 & 0.15 & 13.20 & 0.02 &-0.10 &- &- &3,3 \\
 19 &   WD 0348+339       &DA4   & 38.5 & 16.1 & -0.5 & 12.27 & 0.10 & 12.59 &-0.30 &-0.17 &- &- &3,3 \\
 20 &    HG7-85           &DA    & 38.4 &  9.2 & -1.2 & 12.02 & 0.16 & 12.50 & 0.09 &-0.27 &y &- &0,0 \\
 21 &   WD 0433+270       &DC8   & 20.1 & 26.8 & -0.5 & 14.35 & 0.65 & 13.08 & 0.46 & 1.02 &y &- &3,3 \\
 22 &   WD 0437+122       &DA    & 66.9 & 21.3 &  1.6 & 13.57 & 0.19 &   --- &  --- &  --- &- &- &9,9 \\ 
\hline 
 23 &   WD 0625+415       &DA3   & 47.7 & 28.9 & -2.9 & 11.60 &-0.03 & 11.95 &-0.17 &-0.20 &- &- &3,3 \\
 24 &   WD 0637+477       &DA3.6 & 36.2 & 30.6 & -1.4 & 12.00 & 0.13 & 12.25 &-0.18 &-0.06 &- &- &3,3 \\
 25 &   WD 0641+438       &DA    & 41.9 & 30.2 &  3.0 & 12.42 & 0.06 & 12.66 &-0.09 &-0.03 &- &- &8,4 \\
 26 &   WD 0743+442       &DA5   & 33.5 & 35.6 &  1.3 & 12.24 & 0.10 & 12.60 & 0.05 &-0.14 &- &- &3,3 \\
 27 &   WD 0816+376       &DA5   & 38.7 & 39.6 & -0.9 & 12.80 & 0.22 & 12.70 &-1.18 &-0.10 &- &- &3,3 \\
\hline\hline                                   
 28 &   WD 0233-083.1     &DA    & 53.3 & 32.5 &  1.5 & 16.13 & 0.19 &   --- &  --- &  --- &- &- &x,x \\
 29 &   WD 0300-083.1     &DA4.4 & 37.9 & 25.0 & -0.0 & 15.93 & 0.07 &   --- &  --- &  --- &- &- &x,x \\
 30 &   LP 652-0342       &---   & 31.6 & 25.2 &  0.2 & 14.25 & 0.06 & 14.27 &-0.16 & 0.15 &- &- &5,5 \\
 31 &   WD 0533+322       &DA4   & 11.2 & 36.1 &  1.1 & 16.18 & 0.38 &   --- &  --- &  --- &- &- &1,1 \\
 32 &   WD 0543+436       &DA5   & 21.3 & 30.3 & -2.0 & 15.45 & 0.14 & 14.76 & 1.01 & 0.30 &- &- &3,3 \\
 33 &   WD 0557+237       &DA6   & 13.2 & 34.4 & -2.1 & 16.29 & 0.35 & 15.80 & 0.48 & 0.52 &- &- &2,2 \\
 34 &1RXSJ062052.2+132436 &DA    & 29.7 & 24.1 & -0.6 & 12.94 &-0.30 & 13.33 & 0.09 &-0.51 &y &- &2,2 \\
 35 &   WD 0758+208       &DA    & 38.7 & 36.8 & -0.8 & 17.73 & 0.33 &   --- &  --- &  --- &- &- &x,x \\
 36 &   WD 0816+387       &DA6.5 & 19.1 & 38.6 &  3.8 & 15.16 & 0.30 & 14.67 & 0.49 & 0.52 &- &- &3,3 \\
 37 &   WD 0820+250.1     &DA1.5 & 28.0 & 38.1 &  3.1 & 13.68 &-0.15 & 14.20 &-0.08 &-0.42 &y &- &7,7 \\

\end{tabular}
\tablefoot{Stars No. 1 to 10 are the 10 ''classical'' Hyades white
dwarfs as given, e.g. by \citet{1998AJ....115.1536V}. Stars No. 11 to 22 are probably former Hyades
white dwarfs that fulfil the kinematic and photometric criteria. Stars No. 23 to 27 do
also fulfil the kinematic and photometric criteria, but we rate them as possible non-Hyades 
because of their long distance from the centre in the $Z$ direction (as was done for MS stars in Paper I).
Stars No. 28 to 37 are probable non-Hyades white dwarfs that photometrically do not share
the loci of white dwarfs in the colour-magnitude diagrams. The printed table gives the entries
necessary for the figures in this paper; the full table with additional information
is available from the CDS, Strasbourg, France.}
\tablebib{
(0)~\citet{1995ApJ...448..683S};
(1)~\citet{2001KFNT...17..409K};
(2)~\citet{1999ApJS..121....1M};
(3)~\citet{1994cmud.book.....M};
(4)~\citet{2005AJ....129.2428S};
(5)~\citet{2003ApJ...582.1011S};
(6)~\citet{1987AJ.....94..501K};
(7)~\citet{2008AJ....136..735L};
(8)~\citet{2004AAS...205.4815Z};
(9)~\citet{1992MNRAS.257..257R};
(x)~converted from SDSS ugriz.
}
\end{table*}
In Table \ref{table:1} we summarise the data for the white dwarfs discussed in this paper. Column 1 is a running number, 
column 2 the name(s) of the star in the SIMBAD database, column 3 the spectral type taken from 
\citet{1999ApJS..121....1M}. In column 4 we present the distance D of the star from the Sun as calculated from
the convergent point method, whereas column 5 gives the distance r$_c$ from the cluster centre. Column 6 is the
tangential velocity v$_\perp$ perpendicular to the direction to the convergent point. It is a measure of how well the
motion of the star and cluster coincide. Columns 7 to 11 give $M_V$, $B-V$, $M_J$, $J-K_s$ and $r'-J$. Column 12
describes whether or not the star is detected as an x-ray source (in VizieR), column  13 whether it is a known spectroscopic binary.
Finally, column 14 presents the sources of the $B$ and $V$ magnitudes. For code x,x
we used the transformations from ugriz (SDSS) to $B$, $V$ as given in \citet{2005AJ....130..873J}.
An extended version of Table~\ref{table:1} is published
only in machine-readable form via the CDS. It contains additional entries for each star
including, e.g. precise positions, proper motions, apparent magnitudes to ease the
preparation of follow-up observations, as well as the velocities derived
from the convergent point method. 

The 37 stars in Table \ref{table:1} are divided into four classes. The first ten stars of class 1 are the ``classical''
Hyades. Stars 11 to 22 form class 2 of new probable Hyades co-movers. The five stars of class 3 fulfil
the kinematic and photometric criteria, but are more than 20 pc away from the centre in $Z$ direction
(perpendicular to the galactic plane).
Stars with these characteristics have been ruled out in Paper I, because all of them had discordant
radial velocities (whenever a radial velocity measurement was available).
Finally, class 4 consists of 10 stars that fulfil the kinematic criteria, but would
be sub-luminous in the CMD if set at their predicted distances.
\begin{table}[h!]
\caption{Predicted and measured radial velocities for stars from Table \ref{table:1}}
\centering                          
\label{table:2}
\begin{tabular}{rrcc@{ }r@{ }l}
\hline\hline
Star   & RV$_{\rm{pr}}$      & RV$_{\rm{S}}$         & RV$_{\rm{V}}$       & Ref. & Comments\\
No.    &$[\rm{km\,s}^{-1}]$&$[\rm{km\,s}^{-1}]$&$[\rm{km\,s}^{-1}]$& S,V   &  \\
\hline
 1     & 34.4           & 23$\pm$10          & 37.4             & 1,2 &binary\\
 2     & 35.4           & 46.3$\pm$4.2       &46.3$\pm$4.2      & 5,5  &  \\
 4     & 38.1           & 37.0$\pm$2       &38.7$\pm$0.3        & 1,6 &binary\\
 7     & 38.9           &-10.4$\pm$9.7    &36.7$\pm$1.5         & 7,3 &cpm, sep. 13\arcsec(?)\\
 8     & 39.5           & 43.5$\pm$4.0       &43.5$\pm$4.0      & 5,5  &  \\
10     & 40.2           &105              &105                  & 8,8 &no redshift correction      \\
11     &  9.9           & 9.8$\pm$8.9        &9.8$\pm$8.9       & 7,7 &cpm, sep. 43\arcsec\\
13     & 19.5           & 5.8$\pm$8.6        &5.8$\pm$8.6       & 7,7 &cpm, sep. 2\arcsec\\
18     & 33.0           & 50.9$\pm$3.3       &49.9$\pm$4.6      & 4,5  &  \\
21     & 37.6           &      -           &36.3                 & -,9 &  \\
25     & 36.1           &-11.7$\pm$8.3    &-11.7$\pm$8.3        & 7,7 &cpm, sep. 143\arcsec \\
36     & 34.6           & 19.8$\pm$6.5       & 19.8$\pm$6.5     & 7,7 &cpm, sep. 34\arcsec  \\
\hline                                   
\end{tabular}
\tablefoot{Radial velocities from literature for the stars from Table~\ref{table:1}.
The first column gives the star number, the second column (RV$_{\rm{pr}}$) the radial velocity predicted by the
convergent point method. In the next three columns we show the radial velocities
found in SIMBAD (RV$_{\rm{S}}$), resp. VizieR (RV$_{\rm{V}}$), and the corresponding references.
The sixth column gives comments.}
\tablebib{
(1)~\citet{WILSON53};
(2)~\citet{2000A&AS..142..217B};
(3)~\citet{1994A&AS..108..603B};
(4)~\citet{2003A&A...400..877P};
(5)~\citet{2006A&A...447..173P};
(6)~\citet{2004A&A...418..989N};
(7)~\citet{2002AJ....124.1118S};
(8)~\citet{1967ApJ...149..283G};
(9)~\citet{2003ApJ...596..477Z}
}
\end{table}

We also checked if measured radial velocities of the 37 candidates
were available to compare them
with the predicted ones from the convergent point method. Only for 12 of
the 37 candidates we found
radial velocities in the literature. For six of them (the stars nos. 1,
4,  7, 8, 11, 21),
the predicted and the measured radial velocities (from at least one source) agree well.
EGGR 42 (10) was assumed to have Hyades
radial velocity by \citet{1967ApJ...149..283G} and this was used to obtain its Einstein redshift
(see also the remark on this  star in the next section). WD~0816+387
(36) was already rejected as a Hyades
member by the photometric criteria, so a disagreement between measured
and predicted radial velocities
is to be expected. On the other hand, the discordant radial velocities for
the stars nos. 2, 13, 18, and 25
require a more detailed discussion.

A reliable determination of space velocities of white dwarfs is a
challenging task. For isolated
white dwarfs the apparent radial velocities must be corrected for
gravitational redshift, which requires the
knowledge of the mass-radius ratios, i.e. quantities that cannot be
observed directly. For stars nos. 2, 8, and
18, \citet{2006A&A...447..173P} determined radial velocities from
high-resolution spectra, whereas spectroscopic
distances and gravitational redshifts were computed from the fundamental
parameters derived by \citet{2001A&A...378..556K}.
The relatively high radial velocity for EGGR 26 (2) by
\citet{2006A&A...447..173P} would reject this star as a Hyades member,
though in numerous studies its membership is found to be confirmed \citep[e.g.,][]
{1992AJ....104.1876W,2009ApJ...696...12D}.
The discrepancy for EGGR 26 may probably be explained by underestimated
uncertainties introduced when deriving the redshift
corrections. The same reason for discrepancy may possibly hold
for LP~653-0026 (18), too: recently, \citet{2009A&A...505..441K}
published an updated version of their catalogue of the fundamental parameters
of white dwarfs where
two different sets of parameters were considered to be equally probable for this
star. A re-calculation of the radial velocities
seems to be reasonable for white dwarfs from
\citet{2006A&A...447..173P}.

For four of our candidates (the  stars nos. 7, 11, 13, 25), radial
velocities were obtained by \citet{2002AJ....124.1118S}
from  line-of-sight velocities of M dwarfs in common proper motion
pairs (cpm), each consisting of an M-dwarf and a white dwarf.
The authors assume that typical separations between the components are
about 1000 AU, such that
orbital motion can be neglected. For star no. 13, where
measured and predicted radial velocities
differ by 1.6$\sigma$, the separation is 2\arcsec, corresponding to
about 50 AU, given a distance of 25 pc.
Here the orbital motion cannot be neglected, and even a small correction
of a few $\rm{km\,s}^{-1}$ could make
the difference between measured and predicted radial
velocities insignificant. On the other hand, if the separation is large in a cpm 
pair, the argument of a common radial velocity
becomes weaker because of an increasing probability
of an unphysical optical pair. This could be the case of
star no.~25, WD~0641+438, where the separation between
the white dwarf and its MS companion reaches 143\arcsec.
The PPMXL lists more or less compatible motions for these stars ($\mu =
139$~mas/yr, $\Theta = 180$~deg;
$\mu = 104$~mas/yr, $\Theta = 183$~deg), respectively. However, the
2MASS colours for the MS star indicate that this star should
be a late K dwarf at a distance of at least 500~pc, which excludes them 
as physical binary.

Finally, the reason for the discrepancy between the predicted and measured 
radial velocity \citep{2002AJ....124.1118S} for HZ~9 (7) seems
to be a misinterpretation. \citet{2002AJ....124.1118S} regarded it as a cpm pair
with another star 13\arcsec away. However, the proper motion of the latter is completely
different (11.4 mas/y) from the proper motion of the white dwarf HZ~9 (115 mas/y). 
On the other hand,
\citet{1987AJ.....94..996S} analysed the radial-velocity curve of HZ~9, confirmed
its binary nature, and determined an M dwarf--white dwarf separation of less than 1~AU and
a radial velocity of 36.7~$\rm{km\,s}^{-1}$ for the system, which agrees well
with the predicted radial velocity.
To summarise the above discussion: none of the 10 classical candidates and of the 17 new
probable former Hyades white dwarfs can be unambiguously discarded on the basis of the
presently measured radial velocities.
Given the problematics of obtaining the true radial motion of
the candidates, we can only encourage
new measurements for which this paper may serve as an input catalogue.
\section{Individual stars}~\label{individual}
\begin{description}
 \item [1 = V471 Tau = WD~0347+171] is a spectroscopic binary K2V+DA \citep{2006MNRAS.367.1699H},
included in our sample of 724 Hyades members from Paper I, a strong X-ray source LX~45 = $229.6 \pm 10.0$. 
LX~45 is the X-ray luminosity from \citet{1995ApJ...448..683S},  derived from the assumption that
the star has a heliocentric distance of 45 pc. Units are $10^{28}$ erg\,s$^{-1}$ (0.1-1.8keV).
Its trigonometric parallax from Hipparcos \citep{2009A&A...497..209V} of 22.7 $\pm$ 1.5 mas 
agrees well with the predicted one.
 \item [2 = EGGR 26 = HZ 4 = WD~0352+096] is considered as a certain member in e.g. \citet{1992AJ....104.1876W} and 
\citet{2009ApJ...696...12D}, though seems just to leave the Hyades (the estimated distance from the cluster
centre is 13.5~pc). A weak X-ray source LX~45 $<$ 1.4.
 \item [  3 = EGGR 29 = LB 227 = WD~0406+169:] \citet{2009ApJ...696...12D} use it in their analysis, whereas 
\citet{1992AJ....104.1876W} declare it to be a non-member. \citet{1992AJ....104.1876W} reject this white dwarf as 
a Hyades member since the mass they derived from the surface gravity sets the star to a distance of 60~pc from the Sun,
with a velocity difference to the adopted cluster velocity higher than 12~$\rm{km\,s}^{-1}$. Using the convergent point
method, we obtain a distance of 52~pc from the Sun, a distance from the cluster centre of 7~pc, and the velocity
difference of less than 0.5~$\rm{km\,s}^{-1}$ for EGGR 29. These results support the assumption of its Hyades membership.
A weak X-ray source LX~45 $<$ 0.9.

 \item [ 4 = HR 1358 = HD 27483 = WD~0418+137:] this system consists of two F6V stars
with orbital period of 3.05 days, and a DA3 white dwarf companion  
\citep{1993AJ....106.1113B}. The MS binary is
included in the sample of the 724 Hyades members from Paper I, X-ray source LX~45 $= 19.9 \pm 2.9$.
Its trigonometric parallax from Hipparcos \citep{2009A&A...497..209V} of 21.1 $\pm$ 0.5 mas
agrees well with the predicted one.

 \item [ 5 = EGGR 36 = VR 7 = WD~0421+162] is considered to be a certain Hyades member in \citet{1992AJ....104.1876W} and
\citet{2009ApJ...696...12D}, a weak X-ray source LX~45 $<$ 1.4.

 \item [ 6 = EGGR 37 = VR 16 = WD~0425+168] is used as a certain Hyades member in \citet{1992AJ....104.1876W} and
\citet{2009ApJ...696...12D}, X-ray source LX~45 = 4.2.

 \item [ 7 = EGGR 38 = HZ 9 = WD~0429+176] is a spectroscopic binary DA2.5+dM, included in the sample of the 724 Hyades members 
from Paper I, X-ray source LX~45 = 2.7.

 \item [ 8 = EGGR 39 = HZ 7 = WD~0431+126] is used as a certain Hyades member in \citet{1992AJ....104.1876W} and
\citet{2009ApJ...696...12D}, a weak X-ray source LX~45 $<$ 0.9.

 \item [ 9 = EGGR 316 = LP 475-242 = WD~0437+138] is accepted as a Hyades member \citep{1998AJ....115.1536V}, though
\citet{1992AJ....104.1876W} do not discuss it, because they are rating the data available for the star as too uncertain.
A weak X-Ray, LX45 $<$ 0.9.

 \item [ 10 = EGGR 42 = HZ 14 = WD~0438+108] is used as a certain Hyades member in \citet{1992AJ....104.1876W} and
\citet{2009ApJ...696...12D}, a weak X-Ray LX45 $<$ 1.0. Its radial velocity has been determined by
\citet{1967ApJ...149..283G}, who give an apparent radial velocity of 105 $\rm{km\,s}^{-1}$. In this case the authors did
not try to determine the Einstein redshift from (M/R), but assumed that it is a Hyades member, and must therefore
have the Hyades radial velocity. 

 \item [ 11 = WD 0120-024] is one of the absolutely faintest white dwarf candidates in our Hyades sample, 
it is not
observed in CMC14. Its predicted and observed radial velocities agree well.

 \item [ 12 =  LP 649-0071:]
at the position of this object we find the MS-star no. 8 ($M = 0.17\, M_\odot$) of our sample from Paper I,
and at the same time
there is the blue object LP 649-0071 ($m_{pg} = 16.9$) from Luyten's White Dwarf Catalogues. The proper motions of the blue Luyten object
and our red object coincide remarkably. We took $B=17.31$ from NOMAD and $V=16.53$ from \citet{2003ApJ...582.1011S}. The red
component has a parallax of 29.13~mas $\pm$ 0.41~mas. 
This object is found as 2XMMi J021352.1-033059 in the
XMM Newton Serendipitous Source Catalogue 2XMMi-DR3,
with a flux of 4.0877$\times 10^{-14}$ mW\,m$^{-2}$ (0.2-12keV). Galex \citep{2007ApJS..173..682M} finds
a faint object with FUV magnitude of 24.0 and NUV magnitude of 23.1 at the position of this star, probably too faint
for a white dwarf with this parallax.
The binary nature of this object as well as its white dwarf nature have to be verified.

 \item [ 13 = WD 0217+375] is a component of a close binary with a separation of 2\arcsec, not resolved in our catalogue.
According to \citet{2005AJ....129.2428S}, the spectral type is M5V+DA. A parallax of 39.8~mas is predicted by the
convergent point method. This agrees well with the parallax of 40~mas given for this star in 
the CNS3 \citep{1991adc..rept.....G}.

 \item [ 14 = WD 0230+343:] 2MASS photometric flags ``ACU''.

 \item [ 15 = LP 246-0014] is a high proper motion ($\mu_{\alpha}\cos \delta = 228.4$~mas/yr, $\mu_{\delta} = -50.8$~mas/yr)
blue star, listed in Luyten's White Dwarf Catalogues. It is located in between two brighter stars. From this
region a strong X-ray emission was measured by ROSAT, but it is not clear which of these objects is an X-ray source.
There are no Galex observations in the area around this star.
The white dwarf nature of LP 246-0014 has to be verified.

 \item [ 16 = WD 0259+378] is one of the white dwarfs most distant to the Sun in our candidate sample. It has a relatively
high residual velocity of 4 $\rm{km\,s}^{-1}$ with respect to the Hyades, which is probably a reason for its ``unusual'' 
location $X, Y = $ -54pc, 32pc in Fig.~\ref{xyz}. Its 2MASS photometry is highly uncertain, especially in the $H$ and
$K_s$ bands, which have photometric flags ``UD''.  

 \item [ 17 = WD 0312+220:] 2MASS photometric flags ``BCU''.	

 \item [ 18 = LP 653-0026 = WD~0339-035:] 2MASS photometric flags  ``ABC''. 

 \item [ 19 = WD 0348+339:] 2MASS photometric flags ``ABD''.

 \item [ 20 = HG7-85 = LP 474-95?] is a white dwarf observed by \citet{2009A&A...505..441K}, HS0400+1451
(Hamburg Schmidt survey).
This star, first mentioned by \citet{1962LowOB...5..257G} as a member of the Hyades, is found in the
Prosser \& Stauffer data base
(presently available from J. Stauffer, priv. comm.). The authors identify it with LP 474-95, a star which
cannot be found in the CDS database. 
It is also not contained in \citet{1999ApJS..121....1M}. 
The proper motions in PPMXL are incorrect.
Therefore, we took the proper motions from \citet{2006A&A...448.1235D}.
The star is also contained in \citet{1995ApJ...448..683S}, which give an X-ray luminosity $<$ 1.1 if a distance
of 45 pc is assumed.

 \item [ 21 = GJ 171.2 B = EGGR 40 = WD 0433+270:]
this star forms a cpm pair with BD+26 730, which is included as no. 461 in Paper I.
The measured trigonometric parallax and radial velocity of BD+26 730 agree well with the
predicted ones. Also, the measured radial velocity of the white dwarf (Table \ref{table:2})
coincides well with the predicted one.
With spectral type DC8, the white dwarf WD 0433+270 is the reddest in our candidate sample. Its possible membership 
in the Hyades is extensively discussed in \citet{2008A&A...477..901C}. We further discuss its
importance for the sample in Sect. \ref{discus}.

 \item [ 22 = LP475-249 = WD 0437+122 = Reid 405] is the most distant white dwarf in our candidate sample, too faint
to be measured in 2MASS and CMC14. A possible membership in the Hyades was discussed by
\citet{1992MNRAS.257..257R}. Owing to a relatively
high velocity with respect to the cluster, this star was excluded as a Hyades member. Based on new proper motions
from PPMXL, the residual velocity v$_\perp$ turns out to be about 1.63~$\rm{km\,s}^{-1}$
which is consistent with the Hyades motion.

 \item [ 23 = WD 0625+415:] this white dwarf has proper motions consistent with Hyades membership. Also, its location
in the $M_V $  vs. $  B-V $ diagram indicates a correct distance predicted by the convergent point method. However, at $z = 11.3$~pc
WD 0625+415 is more than 20~pc above the cluster centre. In Paper~I we found that all stars with $\Delta\,z > 20$~pc should
be rejected as Hyades members. Fig.~\ref{vdisp} gives an additional argument that this star is probably a field white dwarf, though
the radial velocity must be measured to support this assumption.   

 \item [ 24 = WD 0637+477] is probably a field white dwarf, the same case as star no. 23 above. 

 \item [ 25 = WD 0641+438] is probably a field white dwarf, the same case as star no. 23 above.

 \item [ 26 = WD 0743+442] is probably a field white dwarf, the same case as star no. 23 above.

 \item [ 27 = WD 0816+376] is probably a field white dwarf, the same case as star no. 23 above.

 \item [ 28 = WD 0233-083.1] is rejected as a Hyades candidate because of its location in the $M_V $  vs. $  B-V $ diagram. The predicted 
distance seems to be underestimated. No 2MASS, CMC14 and reliable $B, V$ measurements are found. $M_{V}$ and $B-V$ are
estimated from SDSS ugriz.

 \item [ 29 = WD 0300-083.1:] the same case as star no. 28 above.

 \item [ 30 = LP 652-0342] is rejected as a Hyades candidate because of its location in the $M_V $  vs. $  B-V $ diagram.
The predicted distance is underestimated. This is supported  by the trigonometric parallax (3.9 $\pm$ 4.2 mas)
from \citet{1995gcts.book.....V}.

 \item [ 31 = WD 0533+322] is rejected as a Hyades candidate because of its location in the $M_V $  vs. $  B-V $ diagram.
The predicted distance seems to be underestimated. 
No 2MASS, CMC14 measurements.

 \item [ 32 = WD 0543+436] is rejected as a Hyades candidate because of its location in the $M_V $  vs. $  B-V $ diagram.
The predicted distance is underestimated.

 \item [ 33 = WD 0557+237:] the same case as star no.32 above.

 \item [ 34 = 1RXSJ062052.2+132436:] the same case as star no.32 above.

 \item [ 35 = WD 0758+208:] the same case as star no.28 above.

 \item [ 36 = WD 0816+387:] the same case as star no.32 above.

 \item [ 37 = WD 0820+250:] the same case as star no.32 above.\\
\end{description}
\noindent Other stars:
\begin{description}
 \item [vA54 = HG7-128 = LP 474-185]
is an M5V star \citep{2010yCat....102023S}, has
Rosat observation (LX45 $<$ 2.2). This is star number 170 of Paper I.
van Altena finds in his first paper \citep{1966AJ.....71..482V} $  B-V $ = 1.82, later 
he corrects it to $  B-V $ = 0.90 \citep{1969AJ.....74....2V}. This is definitely a red dwarf, and there is no indication for a white dwarf companion.
 \item [ vA71 = EGGR 32 = WD 0412+14]
is classified as sdK: by  \citet{2010yCat....102023S}. \citet{1975ApJ...200L..95L} rates it as a very metal-poor
subdwarf with ``K-star''
colour, but with strong, sharp hydrogen lines. The General Catalog of Trigonometric Parallaxes
\citep{1995gcts.book.....V}
gives a parallax of 0.003 $\pm$ 0.004 arcseconds.
This star is not a member of the Hyades. No X-ray detection.
 \item [ RHya 102 = HG7-126:]
this star fails the kinematic criterion to be included as Hyades member. Its tangential velocity v$_\perp$ is
$-6.1~\rm{km\,s}^{-1}$. The convergent point method puts it at a distance D from the Sun of 60.8 pc.
With this distance it has $M_V $ = 12.26; and with $  B-V  = -0.08$ it perfectly fits the CMD in Fig.~\ref{cmd}.
It also fits the luminosity-distance relation in Fig.~\ref{vdisp}, see Sect. \ref{discus}.
We discard it, however, because of kinematic reasons.
 \item [ RHya 154:]
the proper motions given by \citet{1992MNRAS.257..257R}, (76,$-34$) mas/y and those from
the CU subset (67, $-46$) disagree. In consequence the v$_\perp$-component of the tangential velocity
is $-3.1~\rm{km\,s}^{-1}$ for Reid, and $-8.1~\rm{km\,s}^{-1}$ in the CU. One would count it as a kinematic 
member with Reid's data, and discard it with ours. The distance from the Sun is D = 66 pc (Reid), 70.6 pc (CU).
Reid gives $V$ = 16.51 and $  B-V  = -0.22$, which converts into $M_V $ = 12.26, and puts it 1~mag below the white dwarf
sequence. So, we count it as a non-member.
\end{description}
\section{Spatial distribution and completeness}~\label{complete}
\begin{figure}[h!]
   \centering
   \includegraphics[bb=47 109 564 640,angle=270,width=9.0cm,clip]{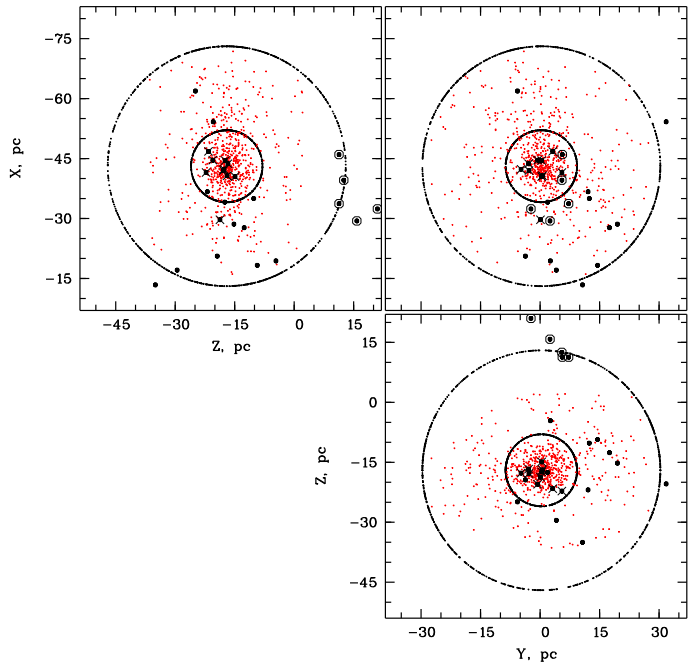}
      \caption{The spatial distribution in the galactic $X, Y, Z$
      coordinate system of the Hyades white dwarf candidates of this paper.  They
      are marked by the same symbols as in Fig. \ref{cmd}. The small (red) dots in the
      background distribution represent the 724 stars from Paper I. Probable field white dwarfs (nos. 23 to 27)
       are additionally
      marked by open circles. The two large circles display the tidal radius (9~pc), and a radius of 30~pc, as in
      Paper I.
   }
         \label{xyz}
   \end{figure}

\begin{figure}[h!]
   \centering
   \includegraphics[bb=53 56 550 585,angle=270,width=9cm,clip]{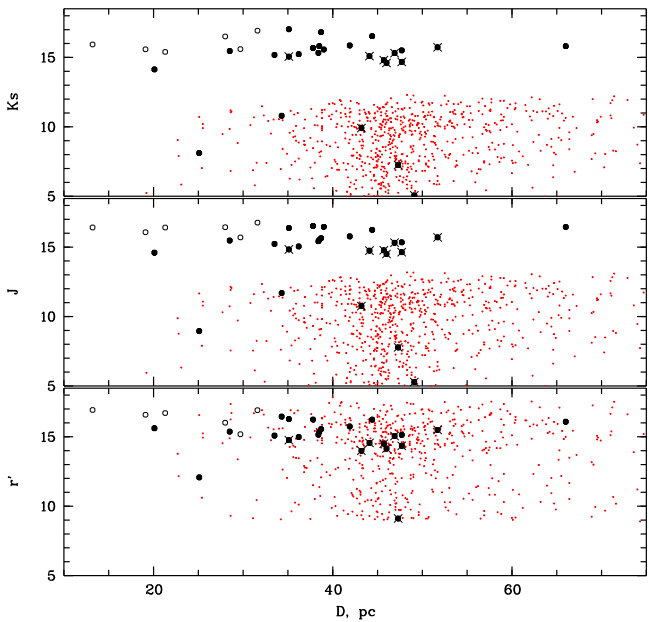}
      \caption{Distribution of the apparent magnitudes $r'$, $J$ and $ K_s$ over
      the distance D from the Sun. The stars from Table~\ref{table:1} 
      are marked by the same symbols as in Fig. \ref{cmd}. The small (red) dots in the
      background distribution represent the 724 stars from Paper I.
   }
         \label{photlim}
   \end{figure}

In Fig. \ref{xyz} we show the distribution of the candidates from Table \ref{table:1} (only stars no. 1 to 27)
on top of the background
of the 724 members from Paper I.
We use the galactic
rectangular coordinate system $X, Y, Z$ with origin in the Sun, and axes pointing to the
Galactic Centre ($X$), to the direction of galactic rotation ($Y$), and to the North 
Galactic Pole ($Z$). 
All classical white dwarfs except one are located within the tidal radius of the cluster.
All newly found candidates lie outside the tidal radius, hence they are no longer gravitationally
bound to the cluster, but share the fate of hundreds of former main-sequence members that
left the bound region. The five probable field white dwarfs (Nos. 23 to 27) are all at z~$>$~10~pc.
Of particular interest here is the distribution in the $X,Y$-plane.
We note that all white dwarfs (except no. 16) follow the tilted distribution of the main-sequence stars.
By tidal interaction with the gravitational field of the Galaxy, stars can leave the cluster on both sides
via the Lagrangian points L$_1$ and L$_2$ of the Galaxy-cluster-star system, where L$_1$ is in the direction
to the Galactic centre, i.e. towards larger (less negative)  $X$, the Sun-facing side of the cluster, while L$_2$
lies on the opposite side of the
cluster centre. All white dwarfs outside the tidal radius (except nos. 16 and 22) populate the Sun-facing part of the cluster, and may have left it
through  L$_1$.
The deficit of newly found candidates at longer distances from the Sun needs explanation. To investigate this
we plot in Fig.~\ref{photlim} the $r'$, $J$ and $ K_s$ magnitudes
as a function of the distance D from the Sun. The background points in Fig.~\ref{photlim} represent again
the sample of 724 from Paper I. In the NIR distributions we note that the magnitude limit of the sample of 724
in the $J$ and $ K_s$ bands is at much brighter magnitudes than the 2MASS completeness limit of $J$ = 15.8 and
$ K_s$ = 14.3 
(see http://www.ipac.caltech.edu/2mass/releases/allsky/doc). For fainter red dwarfs or even brown dwarfs there
was no optical counterpart in CMC14, i.e. in the CU subset to PPMXL.
This is different with the white dwarfs. We see from Fig.~\ref{photlim} that the fainter,
hitherto unknown Hyades white dwarfs all are well beyond the 2MASS completeness limit in $J$ and $ K_s$. The 
photometric accuracy in the $ K_s$ band at 16.0 typically is 0.25 mag, fainter ones have no
accuracy estimate at all. The situation is somewhat better
in the $J$ band. In the $r'$ band the red and white dwarfs are comparable. The faintest
white dwarfs are near the completeness limit of CMC14 at $r'$ = 16.8. With these remarks
it becomes clear that we can reveal new Hyades white dwarfs beyond a distance
of about 50 pc from the Sun
in the CU subset of PPMXL only by chance.
  
The PPMXL goes about 3 magnitudes deeper
than its CU subset, so possibly white dwarfs with Hyades motion could be found therein.
However, PPMXL photometry in optical bands is from USNO-B1.0, and therefore inappropriate for this kind of work.
These distant white dwarfs  could only be found by cross-matching
PPMXL kinematic candidates with the catalogue from \citet{1999ApJS..121....1M}
if the latter would be complete down to the limiting magnitude of PPMXL.
However, except for its SDSS part, which only marginally
covers the Hyades area, the McCook\&Sion catalogue is quite incomplete already at $V \approx$ 16 \citep[see,
e.g.][]{2003Msngr.112...25N}.
Consequently,
our Table~\ref{table:1} contains only two white dwarfs farther away than 50 pc from the Sun.

\section{Discussion}~\label{discus}
\begin{figure}[h!]
   \centering
   \includegraphics[bb=40 52 480 506,angle=270,width=8cm,clip]{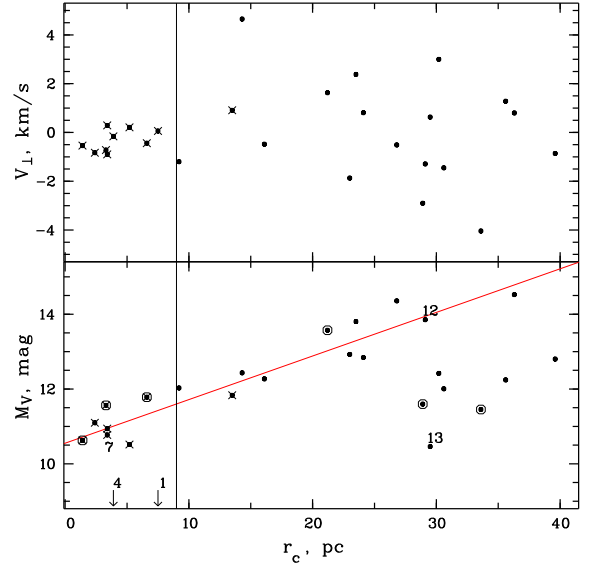}
      \caption{As a function of the distance
      from the cluster centre (r$_c$) this figure shows
      the distribution of v$_\perp$, the velocity component perpendicular
      to the direction to the convergent point (upper panel);
      the absolute magnitudes $M_V $ derived from the secular
      parallaxes (lower panel). White dwarf
      candidates with distances D $>$ 46.3 pc from the Sun are additionally
      marked by larger circles. Spectroscopic binaries are marked by their numbers in Table~\ref{table:1}.
      The red line shows an approximate fit to an
      empirical magnitude-distance
      relation explained in the text.}
         \label{vdisp}
   \end{figure}

The convergent point method supplemented by photometric selection provides five out of six phase-space
parameters. Generally, this allows quite a reliable selection of open cluster members.
We find that only nine ``classical'' Hyades white dwarfs
reside within the tidal radius of the cluster (r$_c < 9$~pc), hence are tidally bound.
Outside the tidal radius, up to a distance of 40 pc from the centre, we find
18 white dwarfs co-moving with the cluster at relatively low velocity
dispersion (cf. Fig.~\ref{vdisp}, top panel).

When \citet{2008A&A...477..901C} discussed WD 0433+270 (no.~21 in Table~\ref{table:1}), this star
was very isolated in the $M_V $  vs. $  B-V $ colour-magnitude diagram of Fig.
\ref{cmd}. There was a large gap between the reddest classical white dwarf
EGGR 26 (no.~2) at $M_V $= 11.83 and $  B-V $= 0.10, and WD 0433+270 at
$M_V $ = 14.35  and $  B-V $ = 0.65. With our new candidates the gap does no longer
exist, and, from kinematics, we strongly infer that WD 0433+270 was in fact a former member of
the Hyades with all the implications that \citet{2008A&A...477..901C} rate as
tantalising. Although the question of the cooling age is most critical for WD 0433+270, if it is an 
ejected member of the Hyades cluster,
the cooling ages of the other candidates between the dimmest classical white dwarf, 
EGGR 26 \citep[$3.1\times10^8$ yr, see][]{1992AJ....104.1876W} 
and WD 0433+270 should be re-discussed after they are confirmed by their radial velocities.
Note that WD 0433+270 is the most nearby star of all our candidates
at a distance of 20.1 pc from the Sun. Given
the incompleteness of the catalogue by \citet{1999ApJS..121....1M}, it may not
be surprising to detect other ``red" white dwarfs of the Hyades once a deep, accurate optical
photometric survey like, e.g. PanSTARSS becomes available.

The one-dimensional velocity dispersion in Fig. \ref{vdisp} increases with increasing
distance from the cluster centre.
This behaviour is similar to that of the red
dwarfs as shown in Fig. 12 of Paper I. From this we infer that the
white dwarfs we reveal here can leave the cluster by the same mechanism as the red dwarfs do.
Once the
progenitor star develops into a white dwarf and its envelope is pushed away,
it will be treated within the cluster as a low-mass object such as the
other low-mass stars that are preferentially ejected from the cluster compared
to their higher mass brothers. This may mean
that, in general, an additional mechanism as proposed by \citet{2003ApJ...595L..53F} is not
needed to explain the white dwarf distribution we find. It is not ruled out that 
the dynamical process of \citet{2003ApJ...595L..53F} has not been active in the Hyades, but
the kilometre-per-second kicks that the stars got would very probably move them
away from the centre much faster, so we would be unable to find most of them with the constraints we adopted.

In Paper I we found mass segregation for giants and main-sequence
stars in the Hyades,
i.e., a strong concentration of the most massive stars ($M >
2\,M_{\odot}$) towards the cluster centre
and flatter distributions for lower mass stars.
Usually, such a concentration of the massive stars is observed already in the
first few $10^7$
yr of a cluster's life (within the relaxation time scale). Since these stars are the
progenitors of white dwarfs, one expects to find recently formed white
dwarfs
in the vicinity of the cluster centre. However, once they are no longer
massive, they behave like other
0.6 to 0.8~$M_{\odot}$ main-sequence stars. Owing to tidal interaction
with the gravitational field
of the Galaxy, the chance of evaporation from the cluster becomes higher,
and it is increasing with
the time passed after degeneration. Therefore, merely from the point of view
of dynamical evolution,
we could expect the older white dwarfs at longer distances from the
cluster centre.
For the white dwarfs of this paper we find that only
the absolutely brightest white dwarfs still are within the tidal radius,
whereas the dimmer ones left the cluster. As we already noted above, the
dimmest of the classical
white dwarfs, EGGR 26, has  a mass of 0.62 $M_\odot$ and a cooling age of
$3.1\times10^8$ yr
\citep{1992AJ....104.1876W}. We find it 13.5 pc away from the cluster
centre, so it
is already outside the tidal radius of the cluster. In the lower panel
of Fig. \ref{vdisp}
we see that the absolute magnitude $M_V $ of white dwarfs increases with
increasing distance from the centre,
to show that possibly the more distant (from the centre) white dwarfs had more time to move
away from the cluster centre and to cool down.
Those must have formed earlier from more massive progenitors. This empirical
luminosity-distance
relation has approximately a slope of 4.5 mag in 40 pc.

In Fig. \ref{vdisp} (bottom)
the five stars marked with their running numbers from Table \ref{table:1}
are binaries, their absolute brightness in $M_V $ is not representative for the white dwarf
component.
However, there is a group of six stars in the bottom panel of Fig.\ref{vdisp} that do not follow
the simple luminosity-distance relation of the others. They lie roughly between 30 and 40 pc
from the centre at $M_V $ between 11.5 and 13.0 mag. Five of these stars (nos. 23 to 27)
are marked as possible field stars because they are far away from the centre in $Z$ direction.
Kinematic and photometric main-sequence candidates have been rejected in Paper I with the same argument.
The loci of stars nos. 23 to 27 in Fig.\ref{vdisp} (bottom) give additional arguments to rule them out and mark them as field white dwarfs.
The sixth star, WD~0259+378 (no. 16), has low  $z-z_{\rm{centre}}$, which is why we keep it as a probable member for 
the time being. We note that these six stars have proper motions and loci in the $M_V$ vs. $B-V$ diagram
that are consistent with Hyades membership. With the radial velocities measured, one will be able to decide on their
membership with more reliability. On the other hand, we should not exclude the possibility that these stars
experienced an additional kick when leaving the cluster, and WD~0259+378 may be a good example of the dynamical
mechanism proposed by \citet{2003ApJ...595L..53F}.

As has been explained in Sect. \ref{complete}, we cannot make a claim for the completeness
of our sample of Hyades white dwarf candidates. However, the probability seems to be low to detect new
white dwarfs of $M_V < 12$ within 10~pc from the cluster centre (or 36~pc $<$ D $<$ 56~pc): the apparent 
magnitudes of these stars would clearly be $V < 16$ where the catalogue of \citet{1999ApJS..121....1M} is nearly
complete in this region.
Indeed, the number of white dwarfs in front of and behind the cluster centre (D = 46.3~pc) is well balanced
within r$_c < 10$~pc. The detection of dimmer white dwarfs is, however, biased towards shorter
distances D from the Sun. If the population of white dwarfs behind the centre were similar
to that in front of it, and the empirical luminosity-distance relation is valid,
we estimate that one will find in the future some 8 to 12  more white dwarfs within
a radius of 40 pc from the centre of the cluster. Depending on whether the white dwarfs (nos. 23 to 27)
are excluded from, or are included in the consideration, this would yield
a total number of 30 to 40 white dwarfs as present-day plus former members of the
Hyades. This  number coincides well with the postulation of \citet{1992AJ....104.1876W},
who claimed that there should be at least 21 white dwarfs dimmer than their seven confirmed Hyades white dwarfs.
   
Although we did not detect new white dwarf candidates within the tidal radius, but only in a 30 pc sphere
around the centre of the Hyades, this result is similar to that of Paper I. There we found that,
at present, 364
main-sequence stars (275 $M_\odot $) are gravitationally bound, and 360 stars
(160 $M_\odot $) are co-moving outside
the present-day tidal radius of the cluster. This is qualitatively consistent (see Fig. 8 in paper I)
with N-body simulations 
of an open cluster comparable to the Hyades \citep{2009A&A...495..807K}. So, we can expect that white dwarfs
are also
subject to cluster evaporation as main-sequence stars are.
As an alternative, \citet{2007A&A...461..957F} proposed that stars outside the tidal radius of the Hyades,
but co-moving in space with the bulk Hyades motion, could be older field stars trapped in 
orbital resonance with the Hyades cluster, a mechanism already described by \citet{1998AJ....115.2384D}.
With the observations we have so far we cannot decide upon the relative efficiency of the two mechanisms, evaporation or capture.
Radial velocity measurements are needed to confirm the co-moving, but cannot distinguish between the two mechanisms either.
Only the determination of the chemical composition (chemical tracking) of the co-moving stars outside
the tidal radius will finally decide about the origin at least for the main-sequence stars. \citet{2008A&A...487..557P}
found empirically that typical open clusters lose between 3 to 14 $M_\odot\,$Myr$^{-1}$ into the field
in the first 260 Myr of their life. So, the 160~$M_\odot $
in the 30~pc volume around the centre can be easily explained with the Hyades lifetime of some 650 Myr.
The capture mechanism must be at least as efficient as that
to compete with evaporation.

To summarise: Within the tidal radius of the Hyades we only find nine ``classical''
bright white dwarfs. It is very improbable that, at present, more white dwarfs brighter than $M_V $ = 12
are tidally bound in the cluster. Outside the tidal radius we find 18 white dwarfs that are co-moving
with the Hyades cluster and  could be former tidally bound members. As a consequence of our
selection process, the sample presented here is incomplete and is essentially restricted to the Sun-facing
part of the cluster. We find five white dwarfs in binary systems, three were already known
as Hyades members, two are new candidates.
Again, this search is incomplete, because we can reveal them only if the white dwarf nature
of one of the components is already known. There is an empirical luminosity-distance (from cluster centre) relation
such that the white dwarfs are dimming by about 1~mag per 10 pc distance from the centre.
Given the spatial incompleteness of our sample, we estimate that some 20 to 30 white dwarfs should co-move with
the bulk Hyades motion in a volume between 9 pc (tidal radius) and 40 pc from the centre. This number is
consistent with an extrapolation of the present day mass function (PDMF) of the cluster (Fig. 10 of Paper I) towards
white dwarf progenitors.
For a full confirmation of the newly found candidates, more measurements of radial velocities are needed.
At present,  none of the 10 classical candidates and of the 17 new
probably former Hyades white dwarfs can be excluded from membership on the basis of the
available measurents of radial velocities.
For white dwarfs,
the measurements of apparent radial velocities must be corrected for gravitational redshift. This correction requires 
determinations of the mass-radius ratio for each object. This, of course, may introduce additional uncertainties
in the determination of the true (kinematic) radial velocity. Once the
candidates are confirmed, theories of white dwarf evolution are
challenged to explain their nature and their origin.
\begin{acknowledgements}
We thank U. Heber for a helpful discussion.
This research has made extensive use of the SIMBAD database,
operated at CDS, Strasbourg, France. We have made use of the NASA/ IPAC Infrared Science Archive,
which is operated by the Jet Propulsion Laboratory, California Institute of Technology,
 under contract with the National Aeronautics and Space Administration.
This publication makes use of data products from the Two Micron All Sky Survey,
which is a joint project of the University of Massachusetts and the
Infrared Processing and Analysis Center/California Institute of Technology,
funded by the National Aeronautics and Space Administration and the National Science Foundation.
\end{acknowledgements}

\end{document}